# ATTENTION-BASED CONVOLUTIONAL NEURAL NETWORK FOR PERFUSION T2-WEIGHTED MR IMAGES PREPROCESSING

**Alkhimova Svitlana**
Associate Professor of the Department of Biomedical Cybernetics, PhD
National Technical University of Ukraine "Igor Sikorsky Kyiv Polytechnic Institute",
Kyiv, Ukraine

**Diumin Oleksii**
Student of the 6th course
National Technical University of Ukraine "Igor Sikorsky Kyiv Polytechnic Institute",
Kyiv, Ukraine

*Abstract. Accurate skull-stripping is crucial preprocessing in dynamic susceptibility contrast-enhanced perfusion magnetic resonance data analysis. The presence of non-brain tissues impacts the perfusion parameters assessment. In this study, we propose different integration strategies for the spatial and channel squeeze and excitation attention mechanism into the baseline U-Net+ResNet neural network architecture to provide automatic skull-striping i.e., Standard scSE, scSE-PRE, scSE-POST, and scSE Identity strategies of plugging of scSE block into the ResNet backbone. We comprehensively investigate the performance of skull-stripping in T2\*-weighted MR images with abnormal brain anatomy. The comparison that utilizing any of the proposed strategies provides the robustness of skull-stripping. However, the scSE-POST integration strategy provides the best result with an average Dice Coefficient of $0.9810 \pm 0.006$.*

**Keywords:** skull-striping, brain, segmentation, region of interest, deep neural network, dynamic susceptibility contrast perfusion, magnetic resonance imaging.

**Introduction**

Nowadays one of the most commonly used perfusion techniques is dynamic susceptibility contrast (DSC) MR imaging [1]. It produces ultrafast T2- or T2\*-weighted sequences of images that allow providing perfusion analysis in cases of oncological diseases (i.e., examining in detail vascular permeability, vessel caliber, tumor cell size, and cytoarchitecture [2, 3]) and ischemic stroke, neurovascular disease, and neurodegenerative disorders [3, 4].

The procedure of skull-striping, also recognized as the segmentation of brain from non-brain tissues or brain extraction, is one of the preprocessing steps in DSC MR data analysis [5]. It is crucial for accurate results of perfusion parameters assessment since the presence of non-brain tissue pixels in analyzed images can lead to visual artifacts in perfusion maps and falsely high or falsely low perfusion parameter values [6, 7].

**Manual, semi-automatic, and automatic procedures of skull-striping**

There are three main approaches to skull-striping procedure: manual delineation of a brain region, semi-automatic, and automatic segmentation of brain from non-brain tissues.

In the case of manual delineation of a brain region, the operator must be a highly trained specialist to identify different anatomical structures and lesions of the brain in MR images. As manual delineation of a brain region is performed by layer-by-layer determining the brain boundaries over the whole volume of MR data of the human head, it is a complex laborious task. It is complicated by several factors such as low image spatial resolution, the absence of intensity standardization, and obscure boundaries of the brain, especially at border pixels lying near areas with abnormal brain anatomy [8]. Additionally, the manual procedure is generally subjective, i.e., obtained results are suffering from considerable intra- and inter-rater variability [9].

Taking into account all of the above, it is desirable to automate the procedure of skull-striping procedure.

In the case of a semi-automatic segmentation of brain from non-brain tissues, the detection of a brain region is assisted by a variety of tools that provides pixel thresholding. An initial threshold value can be provided automatically by the histogram analysis. In most cases, the thresholding results require further correction. The common user interface of such tools is sliders or turning the wheel or holding down the mouse buttons in a predefined mode.

The current automatic skull-striping procedure can be broadly grouped into two groups: intensity-based methods and template-based methods [10, 11].

Intensity-based methods [12-15] provide inaccurate results of segmentation because of overlapping pixel intensities in regions with abnormal brain anatomy and regions which are targeted to be excluded.

Template-based methods suffer from a lack of pre-segmented templates for different age-sex-race-specific patients and different shapes, densities, and locations of the brain lesions [16]. Thus, the proposed methods currently are applicable for the segmentation of healthy subject images [17] or with specific brain lesions [18].

Recently deep learning-based methods have attracted enormous attention in medical image processing and have become the state-of-the-art for segmentation tasks. **The objective of this study** is to give a comprehensive performance comparison of different integration strategies for the attention mechanisms into the baseline U-Net+ResNet neural network architecture for automatic skull-striping in T2*-weighted MR images with abnormal brain anatomy.

**Materials and experiment**

Concurrent spatial and channel squeeze and excitation (scSE) attention mechanism was proposed as attention mechanisms to get an improvement in performance [19]. However, there can be different strategies for the scSE block integration. The Standard scSE block integration strategy is applied right after the final convolutional layer, right before the merging of the skip connection. The scSE-PRE integration strategy consists of plugged scSE block before the first convolutional layer. The scSE-POST integration strategy consists of plugged scSE block after the merging of the skip connection. Finally, the scSE-Identity integration strategy applies the scSE block in the skip connection branch itself, i.e., parallel to the main block.

We implemented the architecture of a deep neural network that combines U-Net [20] and ResNet [21] to provide automatic skull-striping in T2*-weighted MR images with abnormal brain anatomy (Fig. 1.)

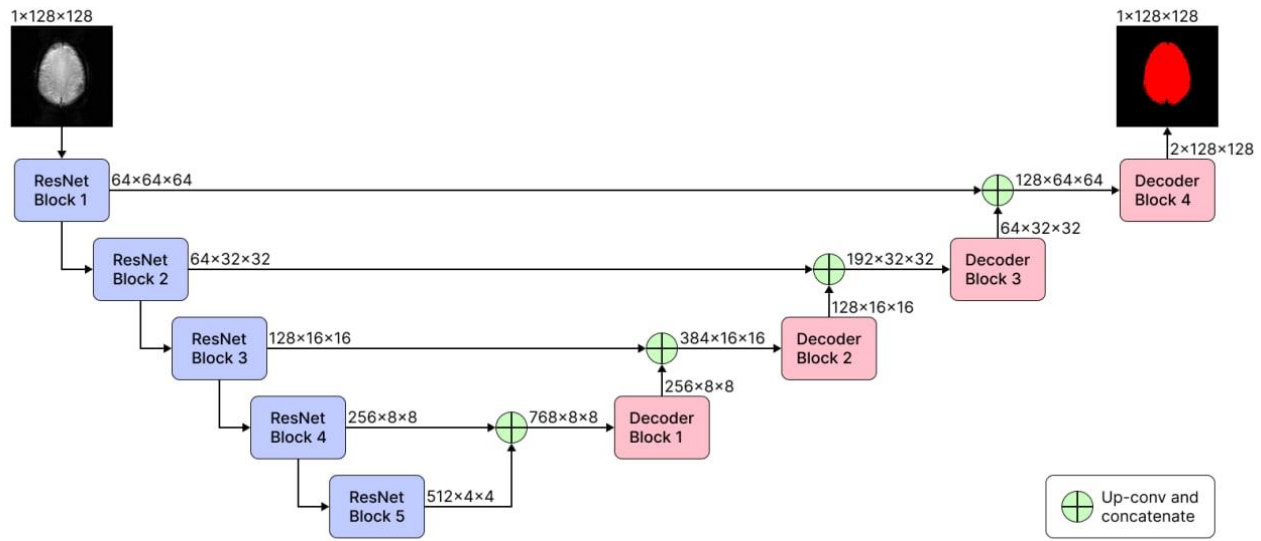

**Figure 1.** The U-Net+ResNet neural network architecture used in this study for skull-striping in T2*-weighted MR images.

To provide a comparison of different integration strategies for the attention mechanisms into the baseline U-Net+ResNet neural network architecture, four strategies of plugging scSE block into the ResNet backbone were implemented, i.e., Standard scSE, scSE-PRE, scSE-POST, and scSE-Identity (Fig. 2).

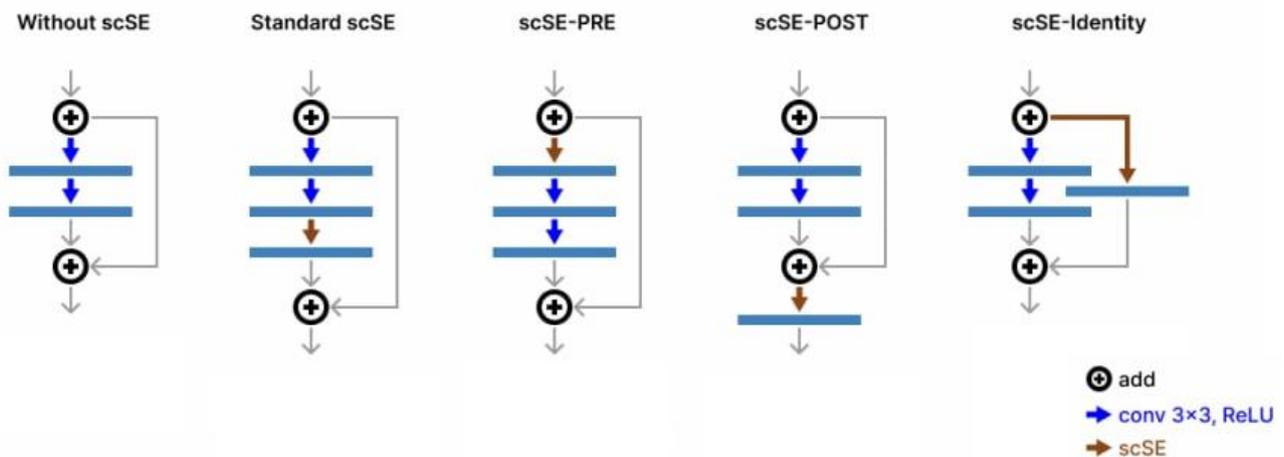

**Figure 2.** scSE block integration designs explored in this study.

We used only T2*-weighted MR data generated by the TCGA Research Network: http://cancergenome.nih.gov/ within a study of Glioblastoma Multiforme. We used 32 three-dimensional volumes of different subjects. The ground-truth masks were manually created by a group of two expert radiologists. Train-validation-test split ratio was 68-12-20. All processing steps were performed using the 4$^{th}$ time-point image from each analyzed space position.

Each model training was performed by a stochastic gradient descent with an Adam optimizer learning rate of 0.00005. In case of no improvement in sparse categorical cross-entropy loss for 10 epochs, the learning rate was divided by 10. The training was run for 100 epochs using mini-batches of 16 images and training data shuffling at the beginning of every epoch.

To quantitatively evaluate the performance of different integration strategies for the scSE attention mechanisms into the baseline U-Net+ResNet architecture, we employ several evaluation metrics such as Dice Coefficient, Sensitivity, Specificity, and Accuracy.

**Results**

We employed the best model of each analyzed case for the test dataset.

The scSE-POST integration strategy of the attention mechanisms offered the best result with an average Dice Coefficient of 0.9810 ± 0.006. The complete evaluation metrics are shown in Table 1.

**Table 1.** Performance comparison of different integration strategies for the scSE attention mechanisms into the baseline U-Net+ResNet architecture, mean ± SD.

|  | Dice Coefficient | Sensitivity | Specificity | Accuracy |
|---|---|---|---|---|
| Without-scSE | 0,9777±0,004 | 0,9630±0,009 | 0,9977±0,001 | 0,9888±0,003 |
| Standard-scSE | 0,9726±0,004 | 0,9514±0,007 | 0,9983±0,001 | 0,9864±0,003 |
| scSE-PRE | 0,9572±0,005 | 0,9243±0,017 | 0,9976±0,002 | 0,9789±0,005 |
| scSE-POST | 0,9810±0,006 | 0,9752±0,009 | 0,9955±0,002 | 0,9904±0,002 |
| scSE-Identity | 0,9621±0,006 | 0,9302±0,010 | 0,9988±0,001 | 0,9813±0,004 |

The loss and accuracy change in the validation phase are plotted in Figure 3. According to the result, it is easy to conclude that scSE-POST integration strategy of the attention mechanisms offers the least loss (0.0363 at 69$^{th}$ epoch) and the best accuracy (0.9855 at 69$^{th}$ epoch).

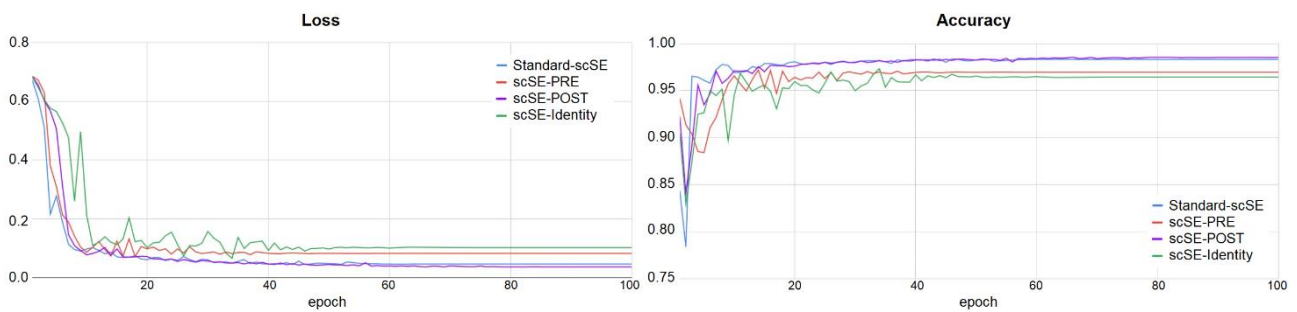

**Figure 3.** The change of loss (left) and accuracy (right) in the validation phase.

The visualization of skull-stripping comparison is shown in Figure 4.

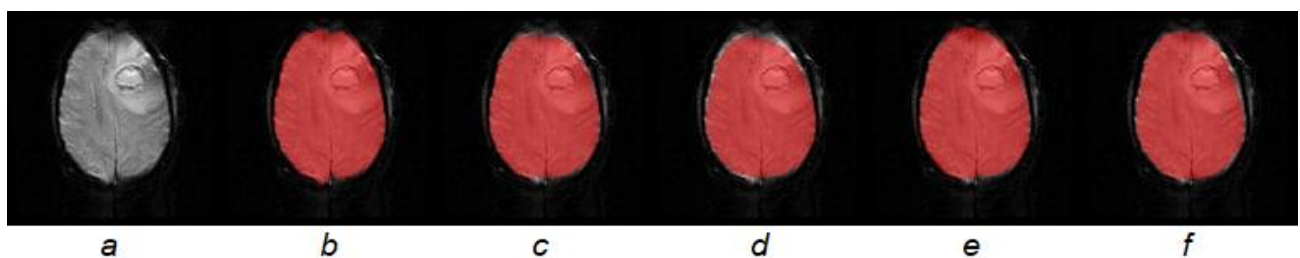

**Figure 4.** Showcases of skull-striping using different scSE block integration strategies: *a* – original image; *b* – ground-truth mask; *c* – Standard scSE; *d* –scSE-PRE; *e* – scSE-POST; *f* – scSE Identity.

The overall performance comparison demonstrates that the scSE-POST integration strategy of the attention mechanisms offers the best results in terms of all evaluation metrics except specificity (i.e., this strategy produces a little bit higher level of missed true pixels of the non-brain region.). The small value of the standard deviation for all integration strategies indicates the robustness of deep learning-based skull-stripping using U-Net+ResNet neural network architecture with scSE attention mechanisms in T2*-weighted MR images with abnormal brain anatomy.

**Conclusions**

In this study, we conduct a performance comparison of different integration strategies for the attention mechanisms into the baseline U-Net+ResNet neural network architecture for automatic skull-striping.

All integration strategies show the robustness of skull-stripping in T2*-weighted MR images with abnormal brain anatomy. However, the scSE-POST integration strategy of the attention mechanisms provides the best result.